\documentclass{article}                  
\usepackage{amsmath} 
\usepackage{amsthm} 
\usepackage{latexsym} 
\usepackage{amssymb}
\usepackage{color}
\usepackage{graphics}

\newtheorem{theorem}{Theorem}
\newtheorem{corollary}{Corollary}
\newtheorem{lemma}{Lemma}
\newtheorem{claim}{Claim}
\newtheorem{proposition}{Proposition}

\theoremstyle{definition}
\newtheorem{definition}{Definition}

\newcommand{\fA}{\ensuremath{\mathfrak{A}}}  
\newcommand{\N}{\ensuremath{\mathbb{N}}}     
\newcommand{\Q}{\ensuremath{\mathbb{Q}}}     
\newcommand{\cC}{\ensuremath{\mathcal{C}}}   
\newcommand{\cE}{\ensuremath{\mathcal{E}}}   
\newcommand{\cL}{\ensuremath{\mathcal{L}}}   
\newcommand{\cM}{\ensuremath{\mathcal{M}}}   
\newcommand{\cN}{\ensuremath{\mathcal{N}}}   
\newcommand{\cS}{\ensuremath{\mathcal{S}}}   

\newcommand{\myvec}[1]{\ensuremath{\mathbf{#1}}}   
\newcommand{\mymat}[1]{\ensuremath{\mathbf{#1}}}   

\newcommand{\kludgeSem}{\ensuremath{\mid \hspace{-0.08cm} \approx}}

\begin{document}
\title{On the Complexity of the Numerically Definite Syllogistic and
  Related Fragments}
\author{Ian Pratt-Hartmann\\
School of Computer Science,\\
University of Manchester,\\
Manchester M13 9PL, U.K.} 
\date{}
\maketitle
\begin{abstract}
\noindent
In this paper, we determine the complexity of the satisfiability
problem for various logics obtained by adding numerical quantifiers,
and other constructions, to the traditional syllogistic. In addition,
we demonstrate the incompleteness of some recently proposed
proof-systems for these logics.

\end{abstract}

\section{Introduction}
\label{sec:intro}
Inspection of the argument
\begin{equation}
\mbox{
\begin{minipage}{10cm}
\begin{tabbing}
{\sf At least 13 artists are beekeepers}\\
{\sf At most 3 beekeepers are carpenters}\\
\underline{\sf At most 4 dentists are not carpenters}\\
{\sf At least 6 artists are not dentists}.
\end{tabbing}
\end{minipage}
}
\label{eq:arg1}
\end{equation}
shows it to be {\em valid}: any circumstance in which all the premises are
true is one in which the conclusion is true. Considerably more thought
shows the argument
\begin{equation}
\mbox{
\begin{minipage}{10cm}
\begin{tabbing}
{\sf At most 1 artist admires at most 7 beekeepers}\\
{\sf At most 2 carpenters admire at most 8 dentists}\\
{\sf At most 3 artists admire at least 7 electricians}\\
{\sf At most 4 beekeepers are not electricians}\\
{\sf At most 5 dentists are not electricians}\\
\underline{\sf At most 1 beekeeper is a dentist}\\
{\sf At most 6 artists are carpenters}
\end{tabbing}
\end{minipage}
}
\label{eq:arg2}
\end{equation}
to be likewise valid---assuming, that is, that the quantified subjects
in these sentences scope over their respective objects. This paper
investigates the computational complexity of determining the validity
of such arguments.

Argument~\eqref{eq:arg1} is couched in a fragment of English obtained
by extending the syllogistic (the language of the syllogism) with
numerical quantifiers. Adapting the terminology of de
Morgan~\cite{cndsrs:deMorgan47}, we call this fragment the {\em
numerically definite syllogistic}.  When its sentences are expressed,
in the obvious way, in first-order logic with counting quantifiers,
the resulting formulas feature only one variable.
Argument~\eqref{eq:arg2} is couched in a fragment of English obtained
by extending the numerically definite syllogistic with transitive
verbs. We call this fragment the {\em numerically definite relational
syllogistic}.  When its sentences are expressed, in the obvious way,
in first-order logic with counting quantifiers, the resulting formulas
feature only two variables.

The satisfiability and finite satisfiability problems for the
two-variable fragment of first-order logic with counting quantifiers
are known to be NEXPTIME-complete. Surprisingly, however, no
corresponding results exist in the literature for the other fragments
just mentioned.  The main results of this paper are: (i) the
satisfiability problem (= finite satisfiability problem) for any logic
between the numerically definite syllogistic and the one-variable
fragment of first-order logic with counting quantifiers is strongly
NP-complete; and (ii) the satisfiability problem and finite
satisfiability problem for any logic between the numerically definite
relational syllogistic and the two-variable fragment of first-order
logic with counting quantifiers are both NEXPTIME-complete, but
perhaps not strongly so. We investigate the related problem of {\em
probabilistic} (propositional) satisfiability, and use the results of
this investigation to demonstrate the incompleteness of some
proof-systems that have been proposed for the numerically definite
syllogistic and related fragments.
\section{Preliminaries}
\label{sec:prelim}
In the sequel, we employ first-order logic extended with the {\em
counting quantifiers} $\exists_{\leq C}$, $\exists_{\geq C}$ and
$\exists_{= C}$, for any $C \geq 0$, under the obvious semantics. Note
that, in this language, $\exists x \phi$ is logically equivalent to
$\exists_{\geq 1} x \phi$, and $\forall x \phi$ is logically
equivalent to $\exists_{\leq 0} x \neg \phi$. The {\em one-variable
fragment with counting quantifiers}, here denoted $\cC^1$, is the set
of function-free first-order formulas featuring at most one variable,
but with counting quantifiers allowed. We assume for simplicity that
all predicates in $\cC^1$ have arity at most 1.

We define the fragment $\cN^1$ to be the set of $\cC^1$-formulas of
the forms
\begin{equation}
\begin{array}{ll}
\exists_{\geq C} x (p(x) \wedge q(x)) \hspace{2cm} &  
\exists_{\geq C} x (p(x) \wedge \neg q(x))\\
\exists_{\leq C} x (p(x) \wedge q(x)) & 
\exists_{\leq C} x (p(x) \wedge \neg q(x)),
\end{array}
\label{eq:formsN1}
\end{equation}
where $p$ and $q$ are unary predicates.  Linguistically, we think of
unary predicates as corresponding to common nouns, and the
formulas~\eqref{eq:formsN1} to English sentences of the forms
\begin{equation}
\begin{array}{ll}
\mbox{{\sf At least} $C$ $p$ {\sf are} $q$} \hspace{2cm} &  
  \mbox{{\sf At least} $C$ $p$ {\sf are not} $q$} \\
\mbox{{\sf At most} $C$ $p$ {\sf are} $q$} &  
  \mbox{{\sf At most} $C$ $p$ {\sf are not} $q$},
\end{array}
\label{eq:formsN1Eng}
\end{equation}
\noindent
respectively. (We have simplified the presentation here by ignoring
the issue of singular/plural agreement; this has no logical or
computational significance, and in the sequel, we silently correct any
resulting grammatical infelicities.)  We call the fragment of English
defined by these sentence-forms the {\em numerically definite
syllogistic}.

The sentence {\sf Some} $p$ {\sf are} $q$ may be equivalently written
{\sf At least} 1 $p$ {\sf is a} $q$, and the sentence {\sf All} $p$
{\sf are} $q$ may be equivalently---if somewhat
unidiomatically---written {\sf At most} 0 $p$ {\sf are not} $q$.
Thus, the numerically definite syllogistic generalizes the ordinary
syllogistic familiar from logic textbooks.  Furthermore, the sentence
{\sf There are at least $C$ $p$} may be equivalently written {\sf At
least} $C$ $p$ {\sf are} $p$; and similarly for {\sf There are at
most} $C$ $p$. Some authors take the sentences {\sf Every $p$ is a
$q$} and {\sf No $p$ is a $q$} to imply that there exists some $p$. We
do not adopt this convention. 

We will have occasion below to extend $\cN^1$ slightly.  Let
$\cN^{1+}$ consist of $\cN^1$ together with the set of
$\cC^1$-formulas of the forms
\begin{equation}
\begin{array}{ll}
\exists_{\geq C} x (\neg p(x) \wedge q(x)) \hspace{2cm} &  
\exists_{\geq C} x (\neg p(x) \wedge \neg q(x))\\
\exists_{\leq C} x (\neg p(x) \wedge q(x)) & 
\exists_{\leq C} x (\neg p(x) \wedge \neg q(x)).
\end{array}
\label{eq:formsN1+}
\end{equation}
These formulas correspond to slightly less natural English sentences
with negated subjects as follows:
\begin{equation*}
\begin{array}{ll}
\mbox{{\sf At least} $C$ {\sf non-}$p$ {\sf are} $q$} \hspace{2cm} &  
  \mbox{{\sf At least} $C$ {\sf non-}$p$ {\sf are not} $q$} \\
\mbox{{\sf At most} $C$ {\sf non-}$p$ {\sf are} $q$} &  
  \mbox{{\sf At most} $C$ {\sf non-}$p$ {\sf are not} $q$}.
\end{array}
\end{equation*}

Turning now to Argument~\eqref{eq:arg2}, we take the {\em two-variable
fragment with counting quantifiers}, here denoted $\cC^2$, to be the
set of function-free first-order formulas featuring at most two
variables, but with counting quantifiers allowed.  We assume for
simplicity that all predicates in $\cC^2$ have arity at most 2.  And
we define the fragment $\cN^2$ to be the set of $\cC^2$-formulas
consisting of $\cN^1$ together with all formulas of the forms
\begin{equation*}
\begin{array}{ll}
\exists_{\geq C} x (p(x) \wedge \exists_{\geq D} y (q(y) \wedge r(x,y))) 
   \hspace{0.3cm} &  
\exists_{\geq C} x (p(x) \wedge \exists_{\leq D} y (q(y) \wedge r(x,y)))\\ 
\exists_{\leq C} x (p(x) \wedge \exists_{\geq D} y (q(y) \wedge r(x,y))) &
\exists_{\leq C} x (p(x) \wedge \exists_{\leq D} y (q(y) \wedge r(x,y))),
\end{array}
\end{equation*}
where $p$ and $q$ are unary predicates, and $r$ is a binary predicate.
Linguistically, we think of
binary predicates as corresponding to transitive verbs, and the
above formulas to English sentences of the forms
\begin{equation}
\begin{array}{ll}
\mbox{{\sf At least} $C$ $p$ $r$ {\sf at least} $D$ $q$} \hspace{0.5cm} &  
\mbox{{\sf At least} $C$ $p$ $r$ {\sf at most} $D$ $q$} \\
\mbox{{\sf At most} $C$ $p$ $r$  {\sf at least} $D$ $q$} &  
\mbox{{\sf At most} $C$ $p$ $r$  {\sf at most} $D$ $q$},
\end{array}
\label{eq:formsN2Eng}
\end{equation}
respectively.  (Again, we ignore the issue of singular and plural
phrases.) Note that the sentence-forms in~\eqref{eq:formsN2Eng} may
exhibit scope ambiguities; we have resolved these by stipulating that
subjects always scope over objects. With this stipulation, we
call the fragment of English defined by the
sentence-forms~\eqref{eq:formsN1Eng} and~\eqref{eq:formsN2Eng} the
{\em numerically definite relational syllogistic}.

We take it as uncontentious that the correspondence
between~\eqref{eq:formsN1} and~\eqref{eq:formsN1Eng} provides a
rational reconstruction of the notion of validity for arguments in the
numerically definite syllogistic: such an argument is valid just in
case the corresponding $\cN^1$-sequent is valid according to the usual
semantics of first-order logic with counting quantifiers. Moreover,
for every $\cN^1$-formula, there is another $\cN^1$-formula logically
equivalent to its negation.  Hence, the notion of validity for
$\cN^1$-sequents is dual to the notion of satisfiability for sets of
$\cN^1$-formulas in the standard way. Similar remarks apply to $\cN^2$
and the numerically definite relational syllogistic.

Let $\cL$ be any logic. The {\em satisfiability problem} for $\cL$ is
the problem of determining whether a given finite set of
$\cL$-formulas is satisfiable (has a model); likewise, the {\em finite
satisfiability problem} for $\cL$ is the problem of determining
whether a given finite set of $\cL$-formulas is finitely satisfiable
(has a finite model).  A logic $\cL$ is said to have the {\em finite
model property} if every finite set of satisfiable $\cL$-formulas is
finitely satisfiable.  Thus, $\cL$ has the finite model property just
in case the satisfiability and finite satisfiability problems for
$\cL$ coincide. As usual, we take the {\em size} of any set $\Phi$ of
$\cL$-formulas to be the number of symbols in $\Phi$, counting each
occurrence of a logical connective or non-logical symbol as
1. (Technically, one is supposed to take into account {\em how many}
non-logical symbols occur in $\Phi$; but for the logics considered
here, this would make no difference.) The computational complexity of
the satisfiability problem and the finite satisfiability problem for
$\cL$ can then be understood in the normal way.  Care is required,
however, when the formulas of $\cL$ contain numerical constituents, as
is the case with the logics considered here.  Under {\em unary
coding}, a positive numerical constituent $C$ is taken to have size
$C$; under {\em binary coding}, by contrast, the same constituent is
taken to have size $\lfloor \log_2 C \rfloor +1$, in recognition of
the fact that $C$ can be encoded as a bit string without leading
zeros.  When giving upper complexity bounds, binary coding is the more
stringent accounting method; when giving lower complexity bounds,
unary coding is. In the sequel, binary coding will be assumed, unless
it is explicitly stated to the contrary.  A problem is sometimes said
to be {\em strongly} NP-complete if it is NP-complete (under binary
coding), and remains NP-hard even under unary coding; and similarly
for other complexity classes.

In a logic with negation, a {\em literal} is an atomic formula or the
negation of an atomic formula; in a logic with negation and
disjunction, a {\em clause} is a disjunction of literals. 

Henceforth, all logarithms have base 2.

\section{Complexity of systems between $\cN^1$ and $\cC^1$} 
\label{sec:np}
In this section, we consider logics containing $\cN^1$ but contained
in $\cC^1$.  
\begin{lemma}
The satisfiability problem for $\cN^1$ is NP-hard, even under unary coding.
\label{lma:NP-hard}
\end{lemma}
\begin{proof}
If $G$ is an undirected graph (no loops or multiple edges), a 3-{\em
colouring} of $G$ is a function $t$ mapping the nodes of $G$ to the
set $\{0,1,2\}$ such that no edge of $G$ joins two nodes mapped to the
same value. We say that $G$ is 3-{\em colourable} if a
3-colouring of $G$ exists. The problem of deciding whether a given graph $G$ is
3-colourable is well-known to be NP-hard. We reduce it to $\cN^1$-satisfiability.

Let the nodes of $G$ be $\{1, \ldots, n \}$. For all $i$ ($1 \leq i
\leq n$) and $k$ ($0 \leq k < 3$), let $p_i^k$ be a fresh unary
predicate. Think of $p_i^k(x)$ as saying: ``$x$ is a colouring of $G$ in
which node $i$ has colour $k$''. Let $\Phi_G$ be the set of
$\cN^1$-formulas consisting of
\begin{align}
& \exists_{\leq 3} x (p(x) \wedge p(x))
\label{eq:numberBound} \\
& \{ \exists_{\leq 0} x (p_i^j(x) \wedge p_i^k(x)) \mid 
                            1 \leq i \leq n,  0 \leq j < k < 3 \} 
\label{eq:coloursDisjoint} \\
& \{ \exists_{\geq 1} x (p_i^k(x) \wedge p(x)) \mid 
                            1 \leq i \leq n,  0 \leq k < 3 \} 
\label{eq:cover} \\
& \{ \exists_{\leq 0} x (p_i^k(x) \wedge p_j^k(x)) \mid 
                           \mbox{$(i,j)$ is an edge of $G$}, 0 \leq k < 3 \} 
\label{eq:properColouring} 
\end{align}
We prove that $\Phi_G$ is satisfiable if and only if $G$ is
3-colourable.

Suppose $\fA \models \Phi_G$.  By~\eqref{eq:numberBound}, $|p^\fA| \leq
3$. Fix any $i$ ($1 \leq i \leq n$).  No $a \in p^\fA$ satisfies any
two of the predicates $p_i^0$, $p_i^1$, $p_i^2$,
by~\eqref{eq:coloursDisjoint}; on the other hand, each of these
predicates is satisfied by at least one element of $p^\fA$,
by~\eqref{eq:cover}; therefore, 
$|p^\fA| = 3$, and
each element $a$ of $p^\fA$ satisfies
exactly one of the predicates $p_i^0$, $p_i^1$, $p_i^2$. Now fix any
$a \in p^\fA$, and, for all $i$ ($1 \leq i \leq n$), define $t_a(i)$
to be the unique $k$ ($1 \leq k < 3$) such that $\fA \models
p_i^k[a]$, by the above argument. The
formulas~\eqref{eq:properColouring} then ensure that $t_a$ defines a
colouring of $G$. Conversely, suppose that $t: \{1, \ldots, n \} \rightarrow
\{0, 1, 2\}$ defines a colouring of $G$.  Let $\fA$ be a structure with
domain $A = \{0, 1, 2 \}$; let all three elements satisfy $p$; and,
for all $k \in A$, let $p_i^k$ be satisfied by the single element $k +
t(i)$ (where the addition is modulo 3).  It is routine to verify that
$\fA \models \Phi_G$.  We note that all numerical subscripts in the
formulas of $\Phi$ are bounded by 3. Thus, NP-hardness remains however
those numerical subscripts are coded.
\end{proof}

So much for the lower complexity bound for $\cN^1$. We now proceed to
establish a matching upper bound for the larger fragment $\cC^1$. The
crucial step in this argument is Lemma~\ref{lma:fewNonZeros1}.  To set
the scene, however, we first recall the following textbook result
(see, e.g.~Paris~\cite{cndsrs:paris94}, Chapter~10).  Denote the set
of {\em non-negative rationals} by $\Q^+$.
\begin{lemma}
Let $\cE$ be a system of $m$ linear equations with rational
coefficients.  If $\cE$ has a solution over $\Q^+$, then $\cE$ has a
solution over $\Q^+$ with at most $m$ non-zero entries.
\label{lma:fewNonZeros3}
\end{lemma}
\begin{proof}
We can write $\cE$ as $\mymat{A}\myvec{x} = \myvec{c}$, where
$\mymat{A}$ is a rational matrix with $m$ rows and, say, $L$,
columns, and $\myvec{c}$ is a rational column vector of length
$m$. If $\myvec{b}$ is any solution of $\cE$ in $\Q^+$ with $k > m$
non-zero entries, the $k$ columns of $\mymat{A}$ corresponding to
these non-zero entries must be linearly dependent.  Thus, there exists
a non-zero rational vector $\myvec{b}'$ with zero-entries wherever
$\myvec{b}$ has zero-entries, such that $\mymat{A}\myvec{b}' = \myvec{0}$.
But then it is easy to find a rational number $\varepsilon$ such that
$\myvec{b} + \varepsilon \myvec{b}'$ is a solution of $\cE$ in $\Q^+$
with fewer than $k$ non-zero entries.
\end{proof}
The question naturally arises as to the corresponding bound when
solutions are sought in $\N$, rather than $\Q^+$. Here, the argument
of Lemma~\ref{lma:fewNonZeros3} no longer works, and the bound of $m$
must be relaxed.
\begin{definition}
A {\em Boolean equation} is any equation of the form $a_1 x_1 + \cdots
a_n x_n = c$, where each $a_i$ ($1 \leq i \leq n$) is either 0 or
1, and $c$ is a natural number.
\label{def:booleanEq}
\end{definition}
\noindent
\begin{lemma}
Let $\cE$ be a system of $m$ Boolean equations in $L$ variables.  If
$\cE$ has a solution over $\N$, then $\cE$ has a solution over $\N$
with at most $m \log(L +1)$ non-zero entries.
\label{lma:fewNonZeros1}
\end{lemma}
\begin{proof}
We write $\cE$ as $\mymat{A}\myvec{x} = \myvec{c}$, where $\mymat{A}$
is a matrix of 0s and 1s with $m$ rows and $L$ columns, $\myvec{c}$ is
a column vector over $\N$ of length $m$, and $\myvec{x} = (x_1, \ldots
x_L)^T$. If $\cE$ has a solution over $\N$, let $\myvec{b} = (b_1,
\ldots, b_L)^T$ be such a solution with a minimal number $k$ of
non-zero entries. We show that
\begin{equation}
k \leq m\log(L+1).
\label{eq:Lineq}
\end{equation}
This condition is trivially satisfied if $k = 0$, so
assume $k >0$. Furthermore, by renumbering the
variables if necessary, we may assume without loss of
generality that $b_j > 0$ for all $j$ ($1 \leq j \leq k$).  Now, if $I
\subseteq \{ 1, \ldots, k \}$, define $\myvec{v}_I$ to be the
$m$-element column vector $(v_1, \ldots, v_m)^T$, where
\begin{equation*}
v_i = \sum_{j \in I} \mymat{A}_{i,j}.
\end{equation*}
That is, $\myvec{v}_I$ is the sum of those columns of $\mymat{A}$
indexed by elements of $I$.  Since each $v_i$ ($1 \leq i \leq m$) is a
natural number satisfying
\begin{equation}
v_i \leq L,
\label{eq:sloppy}
\end{equation}
the number of vectors $\myvec{v}_I$ (as $I$ varies over subsets of
$\{1, \ldots, k\}$) is certainly bounded by $(L+1)^m$.  So suppose,
for contradiction, that $k > m\log(L+1)$. Then $2^k > (L+1)^m$, whence
there must exist distinct subsets $I$, $I'$ of $\{1, \ldots, k \}$
such that $\myvec{v}_I = \myvec{v}_{I'}$. Setting $J = I \setminus I'$
and $J' = I' \setminus I$, it is evident that $J$ and $J'$ are
distinct (and disjoint), again with $\myvec{v}_J = \myvec{v}_{J'}$. By
interchanging $J$ and $J'$ if necessary, we may assume that $J \neq
\emptyset$. Now define, for all $j$ ($1 \leq j \leq L$):
\begin{equation*}
b'_j = 
\begin{cases}
b_j -1 \text{ if $j \in J$}\\
b_j +1 \text{ if $j \in J'$}\\
b_j \text{ otherwise,}\\
\end{cases}
\end{equation*}
and write $\myvec{b}' = (b'_1, \ldots, b'_L)^T$.  Since $J$ and $J'$
are disjoint, the cases do not overlap; and since the $b_j$ are all
positive ($1 \leq j \leq k$), the $b'_j$ all lie in $\N$. Moreover,
\begin{equation*}
\mymat{A}\myvec{b}' = \mymat{A}\myvec{b} - \myvec{v}_J + \myvec{v}_{J'}
 =                   \mymat{A}\myvec{b}.
\end{equation*}
Since $J$ is nonempty, $\min \{b'_j | j \in J \}$, is strictly smaller
than $\min \{b_j | j \in J \}$. Generating $\myvec{b}''$,
$\myvec{b}'''$, etc.~in this way (using the same $J$ and $J'$) will
thus eventually result in a vector---say, $\myvec{b}^*$---with
strictly fewer non-zero entries than $\myvec{b}$, but with
$\mymat{A}\myvec{b}^* = \mymat{A}\myvec{b}$---a contradiction.
\end{proof}

By way of a digression, we strengthen Lemma~\ref{lma:fewNonZeros1} to
obtain a bound which does not depend on $L$.
\begin{proposition}
Let $\cE$ be a system of $m$ Boolean equations.  If $\cE$ has a
solution over $\N$, then $\cE$ has a solution over $\N$ with at most
$\frac{5}{2}m\log m +1$ non-zero entries.
\label{prop:fewNonZeros2}
\end{proposition}
\begin{proof}
The case $m=1$ is trivial: if $\cE$ has a solution, then it has a
solution with at most one non-zero entry. So assume henceforth that
$m>1$.

In the proof of Lemma~\ref{lma:fewNonZeros1}, the
inequality~\eqref{eq:sloppy} can evidently be strengthened to
\begin{equation*}
v_i \leq k.
\end{equation*}
Proceeding exactly as for Lemma~\ref{lma:fewNonZeros1}, we obtain, in
place of~\eqref{eq:Lineq}, the inequality
\begin{equation*}
k \leq m\log(k+1).
\end{equation*}
Hence, for $k$ positive, we have
\begin{equation}
\frac{k}{\log(k+1)} \leq m.
\label{eq:kmineq}
\end{equation}
Now the left-hand side of~\eqref{eq:kmineq} is greater than or equal
to unity, and since the function $x \mapsto x \log x$ is monotone
increasing for $x \geq e^{-1}$, we can apply it to both sides
of~\eqref{eq:kmineq} to obtain
\begin{equation}
k Z(k) \leq m\log m,
\label{eq:gammaineq}
\end{equation}
where, for all $k >0$,
\begin{equation*}
Z(k) = \frac{\log k - \log \log (k+1)}{\log (k+1)}. 
\end{equation*}
It is straightforward to check that $Z$ is monotone increasing on the
positive integers, and that $Z(k) \rightarrow 1$ as $k \rightarrow
\infty$.  (Indeed, for $x >0$, the function $x \mapsto \log
x/\log(x+1)$ is monotone increasing with limit 1 as $x$ tends to
$\infty$; and for $x \geq 2^e -1$, the function $x \mapsto \log \log
(x+1)/\log (x+1)$ is monotone decreasing with limit 0.) 

We may now establish that $k \leq \frac{5}{2}m\log m +1$.  Calculation shows
that $1/Z(7) \approx 2.4542 < \frac{5}{2}$.  Therefore, since $Z$ is monotone
increasing, \eqref{eq:gammaineq} yields, for $k \geq 7$, the
inequalities $k \leq m\log
m/Z(k) \leq m\log m/Z(7) < \frac{5}{2} m\log m$.  Obviously, if
$k \leq 6$, we have $k \leq \frac{5}{2}m\log m +1$, since $m \geq 2$ by assumption.
\end{proof}
\noindent

The proof of Proposition~\ref{prop:fewNonZeros2} actually shows a
little more than advertised: for {\em any} real $c>1$, there exists a
$d$ such that, if $\cE$ is a system of $m$ Boolean equations with a
solution over $\N$, then $\cE$ has a solution over $\N$ with at most
$cm\log m +d$ non-zero entries. (As $c$ approaches unity, the required
value of $d$ given by the above proof quickly becomes astronomical.)
It follows that none of these bounds is optimal, in the sense of being
achieved infinitely often. On the other hand, the next lemma shows
that, for systems of Boolean equations with variables ranging over
$\N$, the bound of $m$ reported in Lemma~\ref{lma:fewNonZeros3} is
definitely not available, a fact which will prove useful in
Section~\ref{sec:syllogisms}.
\begin{lemma}
Fix
$m \geq 6$. Let $\mymat{A}$ be the $m \times (m+1)$-matrix given by  
\begin{equation*}
\mymat{A} = \left(
\begin{array}{lllllll|llllll}
1 & 1 & 1 & 0 & 0 & 0 & 0 & 0 & \ldots & \ & \ & \ & 0 \\
0 & 1 & 1 & 1 & 0 & 0 & 0 & 0 & \ldots & \ & \ & \ & 0 \\
0 & 0 & 1 & 1 & 1 & 0 & 0 & 0 & \ldots & \ & \ & \ & 0 \\
0 & 0 & 0 & 1 & 1 & 1 & 0 & 0 & \ldots & \ & \ & \ & 0 \\
0 & 0 & 0 & 0 & 1 & 1 & 1 & 0 & \ldots & \ & \ & \ & 0 \\
0 & 0 & 0 & 0 & 0 & 1 & 1 & 1 & \ldots & \ & \ & \ & 0 \\
\vdots & \ & \ & \ & \ & \ & \ & \ & \ & \ & \ & \ & \vdots \\
0 & 0 & 0 & 0 & 0 & 0 & 0 & 0 & \ldots & 0 & 1 & 1 & 1 \\
\hline
1 & 1 & 0 & 1 & 0 & 0 & 1 & 0 & \ldots & \ & \ & \ & 0 \\  
\end{array}
\right),
\end{equation*}
in which a pattern of three $1$s is shifted right across the first
$(m-1)$ rows, and the last row contains the seven entries shown on the
left followed by $(m-6)$ $0$s.  Let $\myvec{c}$ be the column vector
of length $m$ given by
\begin{equation*}
\myvec{c} = (3, 3, \ldots, 3, 4)^T
\end{equation*}
consisting of $(m-1)$ $3$s and a single $4$.  Then the unique solution
of the system of Boolean equations $\mymat{A}\myvec{x} = \myvec{c}$
over $\N$ is the column vector $(1, \ldots, 1)^T$ consisting of
$(m+1)$ $1$s.
\label{lma:manyNonZeros}
\end{lemma}
\begin{proof}

Evidently, $\mymat{A}(1, \ldots, 1)^T = \myvec{c}$.  Conversely,
suppose $\myvec{b} = (b_1, \ldots, b_{m+1})^T$ is any solution of
$\mymat{A}\myvec{x} = \myvec{c}$ in $\N$.  From the first row of
$\mymat{A}$, $b_1+b_2+b_3 = 3$, whence $b_1, b_2, b_3$ are either (i)
the integers $0, 0, 3$ in some order, or (ii) the integers $0, 1, 2$
in some order or (iii) the integers $1, 1, 1$. By considering rows 2
to $m-1$ of $\mymat{A}$, it is then easy to see that, in every case,
these three values must recur, in the same order, to the end of the
vector: that is, $\myvec{b}$ must have the form
\begin{equation*}
(b_1, b_2, b_3, b_1, b_2, b_3, b_1, \ldots )^T.
\end{equation*}
From the last row of $\mymat{A}$, then, $3b_1 + b_2 = 4$. Thus, $b_1,
b_2, b_3$ are certainly not $3, 0, 0$, in any order. Suppose, then,
$b_1, b_2, b_3$ are $0, 1, 2$, in some order. If $b_1 = 0$, then $3b_1
+ b_2$ is at most 2; if $b_1 = 1$, then $3b_1 + b_2$ equals either 3
or 5; and if $b_1 = 2$, then $3b_1 + b_2$ is at least 6. Thus, $b_1,
b_2, b_3$ are not $0, 1, 2$, in any order, whence $\myvec{b} = (1,
\ldots, 1)^T$ as required.
\end{proof}
\noindent

Returning to the main business of this section, we have:
\begin{theorem}
The fragment $\cC^1$ has the finite model property. Moreover, the
satisfiability (= finite satisfiability) problem for $\cC^1$ is in NP.
\label{theo:c1}
\end{theorem}
\begin{proof}
If $\phi$, $\psi$ and $\pi$ are $\cC^1$-formulas, denote by
$\phi[\pi/\psi]$ the result of substituting $\pi$ for all occurrences
of $\psi$ (as subformulas) in $\phi$.

It is straightforward to transform any $\cC^1$-formula $\phi$, in
polynomial time, into a closed $\cC^1$-formula $\phi'$ containing no
occurrences of equality, no proposition-letters and no individual constants,
such that $\phi$ is satisfiable over a given domain $A$ if and only if
$\phi'$ is satisfiable over $A$. Indeed, we may further restrict
attention to such $\phi'$ having the form
\begin{equation}
\bigwedge_{1 \leq i \leq m} \exists_{\bowtie_i C_i} x \phi_i, 
\label{eq:normalC1}
\end{equation}
where the symbols $\bowtie_i$ are any of $\{ \leq, \geq, = \}$, and
the $\phi_i$ are quantifier free. For suppose $\phi'$ 
does not have this form: we process $\phi'$ as follows. Choose
any quantified subformula $\psi = \exists_{\bowtie C} x \pi$ with
$\pi$ quantifier-free, and non-deterministically replace $\phi'$ by
either $\phi'[\top/\psi] \wedge \psi$ or $\phi'[\bot/\psi] \wedge \neg
\psi$; then repeat this procedure until all embedded
quantification has been removed. The result will be, modulo trivial
logical equivalences, a formula of the form~\eqref{eq:normalC1}; and
$\phi'$ will be satisfiable over a given domain $A$ if and only if some
formula of the form~\eqref{eq:normalC1} obtained in this way is
satisfiable over $A$. Thus, any polynomial-time non-deterministic
algorithm to check the (finite) satisfiability of formulas of the
form~\eqref{eq:normalC1} easily yields a polynomial-time
non-deterministic algorithm to check the (finite) satisfiability of
$\cC^1$-formulas.

Fix $\phi$ to be of the form~\eqref{eq:normalC1}, then, with no
individual constants, proposition letters or equality. Suppose that
the unary predicates occurring in $\phi$ are $p_1, \ldots, p_l$.  Call
any formula of the form $\pi = \pm p_1(x) \wedge \cdots \wedge \pm
p_l(x)$ a 1-type. Let the 1-types be enumerated in some way as $\pi_1,
\ldots, \pi_L$, where $L = 2^l$. Any structure $\fA$ interpreting the
$p_1, \ldots p_l$ can evidently be characterized, up to isomorphism,
by the sequence of cardinal numbers $(\alpha_1, \ldots \alpha_L)$,
where $\alpha_j$ is the cardinality of the set $\{ a \in A : \fA
\models \pi_j[a] \}$ for all $j$ ($1 \leq j \leq L$). Denote this
sequence by $\alpha(\fA)$.  For all $i$ ($ 1 \leq i \leq m$) and $j$
($ 1 \leq j \leq L$), define
\begin{equation*}
a_{i,j} =
\begin{cases}
1 \text{ if $\models \pi_j \rightarrow \phi_i$}\\
0 \text{ otherwise.}
\end{cases}
\end{equation*}
Interpreting the arithmetic operations involving infinite
cardinals in the expected way, if $\fA \models \phi$, then $\alpha(\fA)$ is
a simultaneous solution of
\begin{equation}
\begin{array}{ccccl}
a_{1,1} x_1 + & \ldots + & a_{1,L} x_L & \bowtie_1 & C_1\\
\vdots        & \        & \vdots      & \vdots    & \vdots \\
a_{m,1} x_1 + & \ldots + & a_{m,L} x_L & \bowtie_m & C_m,
\end{array}
\label{eq:arrayEqs}
\end{equation}
with at least one non-zero value. Conversely, given any solution $\alpha_1, \ldots
\alpha_L$ of~\eqref{eq:arrayEqs} with at least one non-zero value, we
can construct a model $\fA$ of $\phi$ such that $\alpha(\fA) =
(\alpha_1, \ldots \alpha_L)$.  Setting $C = \max \{ C_i \mid 1 \leq i
\leq m \}$, we see that, if $\alpha_1, \ldots \alpha_L$ is a solution
of~\eqref{eq:arrayEqs}, then so is $\beta_1, \ldots \beta_L$, where
$\beta_j = \min(\alpha_j, C)$ for all $j$ ($1 \leq j \leq L$).  
It follows easily that $\cC^1$ has the finite model property.

By Lemma~\ref{lma:fewNonZeros1}, \eqref{eq:arrayEqs} has a solution
over $\N$ if and only if it has a solution in which at most $m \log
(L+1) \leq m(l+1) \leq |\phi|^2$ values are nonzero. (The requirement that the
solution in question contain at least one non-zero value can easily be
accommodated by adding one more inequality, if necessary.) By the
reasoning of the previous paragraph, we may again assume all these
non-zero values to be bounded by $C$. But any such solution can be
written down and checked in a time bounded by a polynomial function of
the size of $\phi$.
\end{proof}
\begin{corollary}
The satisfiability problem (= finite satisfiability problem) for any
logic between $\cN^1$ and $\cC^1$ is strongly NP-complete.
\label{cor:np}
\end{corollary}

It follows that determining the validity of arguments in the
numerically definite syllogistic is a co-NP-complete
problem. Equipping this fragment with relative clauses, for example,
\begin{center}
  {\sf At most 3 artists who are not beekeepers are carpenters},
\end{center}
evidently has no effect on the complexity of determining validity,
since it does not take us outside the fragment $\cC^1$.  Nor indeed
has the addition of proper nouns, for the same reason.  In fact, it is
straightforward to show that the complexity bound for satisfiability
(and finite satisfiability) given in Theorem~\ref{theo:c1} applies to
extensions of $\cC^1$ featuring a large variety of other quantifiers.
The only requirement is that the truth-value of a formula $Qx(\psi_1,
\ldots, \psi_n)$ be expressible as a collection of `linear'
constraints involving the cardinalities (possibly infinite) of Boolean
combinations of the sets of elements satisfying the $\psi_1$, \ldots,
$\psi_n$. In particular, we obtain the same complexity for extensions
of the numerically definite syllogistic featuring such sentences as
\begin{center}
\begin{minipage}{10cm}
{\sf There are more artists than beekeepers}\\
{\sf Most artists are beekeepers}\\
{\sf There are more than 3.7 times as many artists as beekeepers}\\
{\sf There are finitely many carpenters}.
\end{minipage}
\end{center}
The details are routine, and we leave them to the reader to explore.
Extensions of the syllogistic with `proportional' quantifiers are
considered by Peterson~\cite{cndsrs:peterson91,cndsrs:peterson79};
however, no complexity-theoretic analysis is undertaken in those
papers, and certainly no analogues of Lemma~\ref{lma:fewNonZeros1}
or Proposition~\ref{prop:fewNonZeros2} are provided.

We conclude this section with some remarks on a related
problem. Denote by $\cS$ the propositional language (with usual
Boolean connectives) over the countable signature of proposition
letters $p_1, p_2, \ldots$.  A {\em probability assignment} for $\cS$
is a function $P: \cS \rightarrow [0,1]$ satisfying the usual
(Kolmogorov) axioms.  The problem PSAT may now be defined as
follows. Let a list of pairs $(\phi_1, q_1), \ldots, (\phi_m, q_m)$ be
given, where each $\phi_i$ is a clause of $\cS$, and each $q_i$ is a
rational number: decide whether there exists a probability assignment
$P$ for $\cS$ such that
\begin{equation}
  P(\phi_i) = q_i \qquad \text{for all $i$ ($1 \leq i \leq m$)}.
\label{eq:psat}
\end{equation}
The size of any problem instance $(\phi_1, q_1), \ldots, (\phi_m,
q_m)$ is measured in the obvious way, with binary coding of the
$q_i$. By comparing~\eqref{eq:psat} with~\eqref{eq:normalC1}, we see
that the satisfiability problem for $\cC^1$ is, as it were, an
`integral' version of PSAT. The problem $k$-PSAT is the restriction of
PSAT to the case where all the clauses $\phi_i$ have at most $k$
literals.

Georgakopoulos {\em et al.}~\cite{cndsrs:gkp88} show that 2-PSAT is
NP-hard, even under unary coding, and that PSAT is in NP. (Hence,
$k$-PSAT is strongly NP-complete for all $k \geq 2$.)  The proof that
2-PSAT is NP-hard is essentially the same as the proof of
Lemma~\ref{lma:NP-hard}. Moreover, the proof that PSAT is in NP is
similar in structure to the proof of Theorem~\ref{theo:c1}. Suppose we are
given an instence of PSAT in which the $\phi_i$ mention only the
proposition letters $p_1, \ldots, p_l$: the challenge is to show that,
if there exists a probability assignment $P$
satisfying~\eqref{eq:psat}, then there exists one in which the number
of formulas $\pm p_1 \wedge \cdots \wedge \pm p_l$ having non-zero
probability is polynomially bounded as a function of $l$. But this is
easily guaranteed by Lemma~\ref{lma:fewNonZeros3}. By contrast,
Lemma~\ref{lma:fewNonZeros3} does not suffice for the proof of our
Theorem~\ref{theo:c1}, because it does not guarantee the existence of
an {\em integral} solution of the relevant equations---hence the need
for Lemma~\ref{lma:fewNonZeros1} (or
Proposition~\ref{prop:fewNonZeros2}). We return to this matter in
Section~\ref{sec:syllogisms}.

\section{Complexity of systems between $\cN^2$ and $\cC^2$} 
\label{sec:nexp}
We now turn our attention to logics containing $\cN^2$ but contained
in $\cC^2$.  
\begin{lemma}
The fragment $\cN^2$ has the finite model property.
\label{lma:fmp}
\end{lemma}
\begin{proof}
Suppose $\fA \models \Phi$, where $\Phi$ is a set of $\cN^2$-formulas.
If $\phi \in \Phi$ is of the form $\exists_{\geq D} x \psi(x)$, let
$A_\phi$ be a collection of $D$ individuals satisfying $\psi$ in
$\fA$, and let
\begin{equation*}
A_\Phi = \bigcup \{ A_\phi \mid \phi \in \Phi \mbox{ is of the form
         $\exists_{\geq D} x \psi(x)$} \}.
\end{equation*}
As in Theorem~\ref{theo:c1}, let the unary predicates occurring in
$\Phi$ be $p_1, \ldots, p_l$, and let $\pi_1(x), \ldots, \pi_L(x)$ be
all the formulas of the form $\pm p_1(x) \wedge \cdots \wedge \pm
p_l(x)$, enumerated in some way.  Then $A$ is the union of the
pairwise disjoint sets $A_1, \ldots A_L$, where $A_j = \{ a \in
\nolinebreak A \mid \fA \models \pi_j[a] \}$. Let $C$ be the largest
quantifier subscript occurring in $\Phi$. Evidently, for all $j$ ($1
\leq j \leq L$), 
\begin{equation*}
| A_\Phi \cap A_j | \leq |A_\Phi| \leq C | \Phi |,
\end{equation*}
whence we may certainly select a set of elements $A'_j$ such that
$A_\Phi \cap A_j \subseteq A'_j  \subseteq A_j$ and
\begin{equation*}
|A'_j| = \min(|A_j|,C | \Phi |+1).
\end{equation*}
Thus $A' = A'_1 \cup \cdots \cup A'_L$ is finite.  We define a
structure $\fA'$ over $A'$ as follows.  Interpret the unary predicates
so that, for all $j$ ($1 \leq j \leq L$) and all $a' \in A'_j$,
$\fA' \models \pi_j[a']$.  Interpret each binary predicate $r$ in such a way
that, for all $a' \in A'$ and all $j$ ($1 \leq j \leq L$),
\begin{equation*}
| \{ b' \in A'_j : \fA' \models r[a',b'] \} | = 
\min(| \{ b \in A_j : \fA \models r[a',b] \} |, C |\Phi| +1).
\end{equation*}
This is evidently possible. Consider any formula $\theta(x)$ of  either
of the forms
\begin{equation}
\exists_{\leq D} y(q(y) \wedge r(x,y)) \hspace{2cm}
\exists_{\geq D} y(q(y) \wedge r(x,y))
\label{eq:predicate}
\end{equation}
with $D \leq C+1 \leq |\Phi|C+1$. It is immediate from the
construction of $\fA'$ that, for all $a' \in A'$,
\begin{equation}
\fA \models \theta[a'] \Rightarrow \fA' \models \theta[a'].
\label{eq:substructureIf}
\end{equation}
Hence, if the numerical subscript $D$ in $\theta$ satisfies $D \leq C$, we also have:
\begin{equation}
\fA \not \models \theta[a'] \Rightarrow \fA' \not \models \theta[a'].
\label{eq:substructureOnlyIf}
\end{equation}
We show that $\fA' \models \phi$ for all $\phi \in \Phi$. If $\phi$ is
an $\cN^1$-formula, this result is immediate from the construction of $\fA'$.
If $\phi$ has the form $\exists_{\geq D'} x(p(x) \wedge \theta(x))$,
with $\theta(x)$ having one of the forms~\eqref{eq:predicate}, the result follows
from~\eqref{eq:substructureIf} and the fact that $A_\phi \subseteq
A'$.  If $\phi$ has the form $\exists_{\leq D'} x(p(x) \wedge
\theta(x))$, with $\theta(x)$ having one of the forms~\eqref{eq:predicate}, the
result follows from~\eqref{eq:substructureOnlyIf}.
\end{proof}
Inspection of the proof of Lemma~\ref{lma:fmp} shows that the size of
the constructed model is bounded by an exponential function of the
size of $\Phi$. Hence, the satisfiability (= finite satisfiability)
problem for $\cN^2$ is in NEXPTIME. 

It is well-known that the larger fragment $\cC^2$ lacks the finite
model property.  For example, the formula
\begin{equation}
\exists x \forall y \neg r(x,y) \wedge
\forall x \exists y r(y,x) \wedge 
\forall x \exists_{\leq 1} y r(x,y)
\label{eq:noFMP}
\end{equation}
is satisfiable, but not finitely so. Thus, the satisfiability problem
and the finite satisfiability problem for $\cC^2$ do not
coincide. Nevertheless, the following was shown in
Pratt-Hartmann~\cite{cndsrs:ph05}.
\begin{theorem}
The satisfiability problem and the finite satisfiability problem for
$\cC^2$ are both in NEXPTIME.
\label{theo:c2inNEXPTIME}
\end{theorem}
\noindent
This upper bound applies even when counting quantifiers are coded in
binary---a fact which is significant here. In this section, we provide
a matching lower-bound for $\cN^2$, which is slightly surprising given
the latter fragment's expressive limitations.

A {\em tiling system} is a triple $\langle C, H, V \rangle$, where $C$
is a finite set and $H$, $V$ are binary relations on $C$. The elements
of $C$ are referred to as {\em colours}, and the relations $H$ and $V$
as the {\em horizontal} and {\em vertical} constraints,
respectively. For any integer $N$, a {\em tiling} for $\langle C, H, V
\rangle$ of size $N$ is a function $t: \{0, \ldots, N-1\}^2
\rightarrow C$ such that, for all $i, j$ in the range $\{0, \ldots,
N-1 \}$, the pair $\langle t(i,j), t(i+1,j) \rangle$ is in $H$ and the
pair $\langle t(i,j), t(i,j+1) \rangle$ is in $V$, with addition
interpreted modulo $N$. A tiling of size $N$ is to be pictured as a
colouring of an $N \times N$ square grid (with toroidal wrap-around)
by the colours in $C$; the horizontal constraints $H$ thus specify
which colours may appear `to the right of' which other colours; the
vertical constraints $V$ likewise specify which colours may appear
`above' which other colours. By a {\em $C$-sequence}, we simply mean a
sequence $\myvec{i} = i_0, \ldots i_{n-1}$ of elements of $C$ (repeats
allowed).  The $C$-sequence $\myvec{i}$ is an {\em initial
configuration} of a tiling $t$ if $\myvec{i} = t(0,0), \ldots,
t(n-1,0)$.

\begin{theorem}
The satisfiability problem for $\cN^2$ is NEXPTIME-hard.
\label{theo:nt}
\end{theorem}
\begin{proof}
Let $\langle C, H, V \rangle$ be a tiling system and $p$ a
polynomial. For any $C$-sequence $\myvec{i}$ of length $n$, we
construct, in time bounded by a polynomial function of $n$, a set
$\Theta_{\myvec{i}}$ of $\cN^2$-formulas such that
$\Theta_{\myvec{i}}$ is satisfiable if and only if $\langle C, H, V
\rangle$ has a tiling of size $2^{p(n)}$ with initial configuration
$\myvec{i}$. Thus, we may regard $\Theta_{\myvec{i}}$ as an {\em
encoding} of the $C$-sequence $\myvec{i}$ with respect to
the tiling system $\langle C, H, V \rangle$ and the function
$2^{p(n)}$.  The existence of such an encoding suffices to show that
the satisfiability problem for $\cN^2$ is NEXPTIME-hard.

To motivate the technical details, we suppose, provisionally, that
$\langle C, H, V \rangle$ does have a tiling of size $2^{p(n)}$ with
initial configuration $\myvec{i}$, and we construct the encoding
$\Theta_{\myvec{i}}$ in parallel with a structure $\fA$ in which
$\Theta_{\myvec{i}}$ is true. As we do so, we show that, conversely,
if $\Theta_{\myvec{i}}$ is satisfiable, then $\langle C, H, V \rangle$
has a tiling with the required properties. The construction of
$\Theta_{\myvec{i}}$ proceeds in two stages. In the first stage, we
employ familiar techniques to obtain an encoding in an extension of
$\cN^2$. In the second stage, we employ some less familiar methods to
obtain an encoding in $\cN^2$.

\vspace{0.25cm}

\noindent
{\bf First stage: }
For convenience, we set $N = 2^{p(n)}$ and $s = 2(p(n)^2+p(n)+1)$.
Let $A_1$ be the set of pairs of integers $(i, j)$ in the range
$\{0, \ldots, N-1 \}$, and let $A_2$ be the set of pairs of the forms $(i,
\top)$ and $(i, \bot)$, where $i$ is an integer in the range $\{1,
\ldots, s \}$ and $\top$, $\bot$ are any distinct symbols. Evidently,
\begin{eqnarray*}
|A_1| & = & N^2\\
|A_2| & = & 2s.
\end{eqnarray*}
Finally, let $A_3$ be a set disjoint from $A_1$ and $A_2$ satisfying
\begin{eqnarray*}
|A_3| & = & (M-1)N^2,
\end{eqnarray*}
where $M = |C|$.  We refer to $A_1$ as the {\em grid}, $A_2$ as the
{\em notebook}, and $A_3$ as the {\em rubbish dump}. Our structure
$\fA$ will have domain $A = A_1 \cup A_2 \cup A_3$. 

Any natural number $l$ in the range $\{0, \ldots, N-1 \}$ can be
written uniquely as $l = \sum_{i=0}^{p(n)-1} b_i 2^i$, where $b_i \in
\{0,1\}$. We say that $b_i$ is the $i$th {\em digit of} $l$. Thus,
digits are enumerated in order of increasing significance,
starting with the zeroth. Let $q$, $X_0, \ldots, X_{p(n) - 1}$,
$\bar{X}_0, \ldots, \bar{X}_{p(n) - 1}$ be new unary
predicates, interpreted in the structure $\fA$ as follows:
\begin{align*}
q^\fA & = A_1 \\
{{X}_i}^\fA & = \{ (l, m) \in A_1 
                      \mid \mbox{ the $i$th digit of $l$ is $1$} \} \\
{\bar{X}_i}^\fA & = \{ (l, m) \in A_1 
                      \mid \mbox{ the $i$th digit of $l$ is $0$} \}. 
\end{align*}
We may read $X_i$ as ``has an $x$-coordinate whose $i$th digit
is 1'', and $\bar{X}_i$ as ``has an $x$-coordinate whose $i$th
digit is 0''. Then $\fA \models \Theta_{0,X}$, where $\Theta_{0,X}$ is
the set of formulas
\begin{align*}
& \exists_{\leq N^2} x q(x) \\
& \exists_{\geq N^2/2} x X_i(x) \\
& \exists_{\geq N^2/2} x \bar{X}_i(x) \\
& \forall x (X_i(x) \rightarrow q(x)) & & (0 \leq i < p(n)) \\
& \forall x (\bar{X}_i(x) \rightarrow q(x)) & & (0 \leq i < p(n)) \\
& \forall x (X_i(x) \rightarrow \neg \bar{X}_i(x)) & & (0 \leq i < p(n)).
\end{align*}
Conversely, in any model of $\Theta_{0,X}$, exactly $N^2$ elements
satisfy $q$, and the extensions of $X_i$ and $\bar{X}_i$ are
complementary with respect to that collection of elements, for all $i$ ($0
\leq i < p(n)$).

Further, let $X^*_0, \ldots, X^*_{p(n)}$ be new unary predicates,
interpreted in the structure $\fA$ as follows:
\begin{align*}
{X^*_0}^\fA & = A_1  \setminus X_0^\fA  \\
{X^*_i}^\fA & =  ({X_0}^\fA \cap \cdots \cap
                      {X_{i-1}}^\fA) \setminus {X_i}^\fA \qquad (1 \leq i < p(n)) \\
{X^*_{p(n)}}^\fA & = {X_0}^\fA \cap \cdots \cap {X_{p(n)-1}}^\fA.
\end{align*}
Thus, the predicate $X^*_i$ can be read as ``has an $x$-coordinate in
which all digits before the $i$th, but not the $i$th digit itself, are
1''. Finally, for all $i, j$ ($0 \leq i < j < p(n)$), let $X^+_{i,j}$
and $X^-_{i,j}$ be new unary predicates, interpreted in the structure
$\fA$ as follows:
\begin{align*}
{X^+_{i,j}}^\fA & = {X^*_i}^\fA \cap {X_j}^\fA\\
{X^-_{i,j}}^\fA & = {X^*_i}^\fA \setminus {X_j}^\fA.
\end{align*}
Let $\Gamma_X$ be the set of first-order formulas 
$\forall x (q(x) \rightarrow \gamma)$, where $\gamma$ is any of the
following clauses:
\begin{align*}
X^*_i(x) \vee &
\big[ X_i(x) \vee \bigvee_{0 \leq k < i} \bar{X}_k(x) \big]
  & \qquad & (0 \leq i < p(n))\\
X^*_{p(n)}(x)  \vee &
\big[\bigvee_{0 \leq k < p(n)} \bar{X}_k(x) \big]
  & \qquad & \ \\
X^+_{i,j}  \vee &
\big[\bar{X}_j(x) \vee X_i(x) \vee 
                    \bigvee_{0 \leq k < i} \bar{X}_k(x)\big] 
     & \qquad & (0 \leq i < j < p(n))\\
X^-_{i,j} \vee &
\big[X_j(x) \vee X_i(x) \vee 
                    \bigvee_{0 \leq k < i} \bar{X}_k(x)\big] 
     & \qquad & (0 \leq i < j < p(n)).
\end{align*}
(The square brackets are for legibility.)  It is immediate that $\fA
\models \Gamma_X$. The formulas $\Gamma_X$, in effect, establish
sufficient conditions for satisfaction of the predicates $X^*_i$ {\em
etc.}~in terms of the predicates $X_i$ and $\bar{X}_i$.  Warning:
these formulas are not in the fragment $\cN^2$.

Similarly, let $Y_0, \ldots, Y_{p(n)-1}$, $\bar{Y}_0, \ldots,
\bar{Y}_{p(n)-1}$ be new unary predicates, interpreted in the
structure $\fA$ as follows:
\begin{align*}
{Y_i}^\fA & = \{ (l, m) \in A_1 
                      \mid \mbox{ the $i$th digit of $m$ is $1$} \} \\
{\bar{Y}_i}^\fA & = \{ (l, m) \in A_1 
                      \mid \mbox{ the $i$th digit of $m$ is $0$} \}.
\end{align*}
We may read $Y_i$ as ``has a $y$-coordinate whose $i$th digit
is 1'', and $\bar{Y}_i$ as ``has a $y$-coordinate whose $i$th
digit is 0''. Let $\Theta_{0,Y}$ be the set of formulas constructed
analogously to $\Theta_{0,X}$, but with ``$X$'' replaced systematically by
``$Y$''; and let $\Theta_0 = \Theta_{0,X} \cup \Theta_{0,Y}$.
Further, let $Y^*_i$ ($0 \leq i \leq p(n)$), $Y^+_{i,j}$ ($0 \leq i <
j < p(n)$) and $Y^-_{i,j}$ ($0 \leq i < j < p(n)$) be new unary
predicates, interpreted analogously to their $X$-counterparts; let the
formulas $\Gamma_Y$ be constructed analogously to $\Gamma_X$; and let
$\Gamma = \Gamma_X \cup \Gamma_Y$.

We may now impose a toroidal grid structure on $A_1$, with the aid of
a pair of binary predicates $h$ and $v$.  Let $\fA$ interpret $h$ and
$v$ as follows:
\begin{eqnarray*}
h^\fA & = & \{ \langle (l, m),  (l+1, m) \rangle \mid
      0 \leq l < N, \ \  0 \leq m < N \} \\
v^\fA & = & \{ \langle (l,m),  (l,m+1) \rangle  \mid
      0 \leq l < N, \ \  0 \leq m < N \},
\end{eqnarray*}
where the addition is modulo $N$. It is straightforward to check that
$\fA \models \Theta_{1,X} \cup \Theta_{1,Y}$, where $\Theta_{1,X}$ is
the set of $\cN^2$-formulas
\begin{align}
& \forall x (q(x) \rightarrow \exists y (q(y) \wedge h(x,y))) 
\label{eq:Theta3.1}\\
& \forall x (X^*_i(x) \rightarrow \neg \exists y (h(x,y) \wedge \bar{X}_i(y))) 
   & & (0 \leq i < p(n))
\label{eq:Theta3.2}\\
& \forall x (X^*_i(x) \rightarrow \neg \exists y (h(x,y) \wedge X_j(y))) 
   & & (0 \leq j < i \leq p(n)) 
\label{eq:Theta3.3}\\
& \forall x (X^+_{i,j}(x) \rightarrow \neg \exists y (h(x,y) \wedge  \bar{X}_j(y))) 
   & & (0 \leq i < j < p(n)) 
\label{eq:Theta3.4}\\
& \forall x (X^-_{i,j}(x) \rightarrow \neg \exists y (h(x,y) \wedge X_j(y))) 
   & & (0 \leq i < j < p(n)) 
\label{eq:Theta3.5}\\
& \forall x (Y_i(x) \rightarrow \neg \exists y (h(x,y) \wedge  \bar{Y}_i(y))) 
   & & (0 \leq i < p(n))
\label{eq:Theta3.6}\\
& \forall x (\bar{Y}_i(x) \rightarrow \neg \exists y (h(x,y) \wedge Y_i(y))) 
   & & (0 \leq i < p(n)),
\label{eq:Theta3.7} 
\end{align}
and $\Theta_{1,Y}$ is defined analogously, but with ``$X$'' and
``$Y$'' interchanged, and ``$h$'' replaced by ``$v$''. Let $\Theta_1 =
\Theta_{1,X} \cup \Theta_{1,Y}$.  Formula~\eqref{eq:Theta3.1} ensures
that every element $a$ satisfying $q$ is related via $h$ to some other
such element $b$. 
In the presence of $\Theta_0$ and $\Gamma$,
\eqref{eq:Theta3.2}--\eqref{eq:Theta3.5} then ensure that
the `$x$-coordinate' of $b$ is one
greater (modulo $N$) than the `$x$-coordinate' of $a$;
and~\eqref{eq:Theta3.6}--\eqref{eq:Theta3.7} likewise ensure that $a$
and $b$ have the same `$y$-coordinate'. Similar remarks apply, {\em
mutatis mutandis}, to the formulas $\Theta_{1,Y}$.  Since $\Theta_0$
ensures that at most $N^2$ elements satisfy $q$, it follows that, in
any model of $\Theta_0 \cup \Gamma \cup \Theta_1$, the extension of
$q$ contains exactly one element with any given pair of
$(x,y)$-coordinates in the range $\{0, \ldots, N-1 \}$, and moreover
that this collection of elements is organized by the interpretations
of $h$ and $v$ into an $N \times N$ toroidal grid in the expected way.

Having set up our grid, we proceed to colour it. Recall that the
`rubbish dump', $A_3$, is a set containing $(M-1)N^2$ elements, where
$M = |C|$.  Let $C = \{ c_1, \ldots, c_M \}$.  Assuming,
provisionally, that $(C, H, V)$ has a tiling of size $N$ with initial
segment $\myvec{i}= i_0, \ldots i_{n-1}$, choose some such tiling
$t$. For all $k$ ($1 \leq k \leq M$), let $n_k \leq N^2$ be the number
of grid-squares to which $t$ assigns colour $c_k$, and let $B_k$ be a
subset of $A_3$ with cardinality $N^2 - n_k$. From the cardinality of
$A_3$, and the fact that $\sum n_k = N^2$, we may choose the $B_k$ to
be pairwise disjoint; and, in that case, the $B_k$ will together
exactly cover $A_3$.  Now treat the elements of $C$ as new unary
predicates, and set
\begin{eqnarray*}
c_k^\fA & = & \{ a \in A_1 \mid \mbox{$t$ assigns the colour $c_k$ to  $a$} \} 
              \cup B_k,
\end{eqnarray*}
for all $k$ ($1 \leq k \leq M$). Let $o$ be a new unary predicate and
set
\begin{eqnarray*}
o^\fA & = & A_1 \cup A_3.
\end{eqnarray*}
It is simple to check that $\fA \models \Theta_2$, where $\Theta_2$ is
the set of formulas 
\begin{align*}
& \forall x (q(x) \rightarrow o(x)) 
\\
& \exists_{\leq MN^2} x o(x)
\\
& \exists_{\geq N^2} x c_k(x) & & (1 \leq k \leq M)
\\
& \forall x (c_k(x) \rightarrow o(x)) & & (1 \leq k \leq M)
\\
& \forall x (c_k(x) \rightarrow \neg c_{k'}(x)) & & (1 \leq k < k' \leq M).
\end{align*}
Conversely, in any model of $\Theta_2$, the interpretations of the
predicates $c_k$ form a pairwise disjoint cover of the interpretation
of $o$, and, therefore, of the interpretation of $q$.

Turning to the input $\myvec{i}= i_0, \ldots, i_{n-1}$, let $o_0,
\ldots, o_{n-1}$ be new unary predicates. We interpret these so as to
pick out the squares $(0,0), \ldots, (n-1,0)$ of the grid,
respectively. Formally:
\begin{eqnarray*}
o_i^\fA & = & \{(i,0) \}
\end{eqnarray*}
for all $i$ ($0 \leq i < n$). It is a simple matter to write formulas
specifying the coordinates of these predicates. For example, define
$\Theta_{3,0}$ to be the set of formulas
\begin{align*}
&\exists x (o_0(x) \wedge q(x)) \\
&\forall x (o_0(x) \rightarrow \bar{X}_i(x)) & (0 \leq i < p(n))\\
&\forall x (o_0(x) \rightarrow \bar{Y}_i(x)) & (0 \leq i < p(n))\\
&\forall x (o_0(x) \rightarrow i_0(x)).
\end{align*}
(Remember that $i_0$, being an element of $C$, is also a predicate
interpreted by $\fA$.)  It is easy to see that $\fA \models
\Theta_{3,0}$.  Conversely, in any model of $\Gamma \cup \Theta_0 \cup
\Theta_1 \cup \Theta_2 \cup \Theta_{3,0}$, $o_0$ must be interpreted
as the (unique) element in the extension of $q$ with `coordinates'
$(0,0)$; moreover, that element must be assigned the `colour'
$i_0$. Let the sets of formulas $\Theta_{3,1}, \ldots, \Theta_{3,n-1}$
be constructed analogously, fixing the interpretations of $o_1,
\ldots, o_{n-1}$, respectively, with colours assigned as specified in
$\myvec{i}$; and let $\Theta_3$ be $\Theta_{3,0} \cup \cdots \cup
\Theta_{3,n-1}$.

Finally, let $\Theta_4$ be the set of formulas 
\begin{align*}
& \forall x (c_j(x) \rightarrow \neg \exists y (c_k(y) \wedge h(x,y)))
  & & 
(1 \leq j \leq M, \ 1 \leq k \leq M, \ (j,k) \not \in H)\\
& \forall x (c_j(x) \rightarrow \neg \exists y (c_k(y) \wedge v(x,y)))
  & & 
(1 \leq j \leq M, \ 1 \leq k \leq M, \ (j,k) \not \in V).
\end{align*}
Since the interpretations of the $c_k$ were taken from a tiling $t$,
we certainly have $\fA \models \Theta_4$. Conversely, any model of
$\Gamma \cup \Theta_0 \cup \cdots \cup \Theta_4$ defines a
tiling for $(C,H,V)$ of size $N$ with initial segment $\myvec{i}$, in
the obvious way.

The set of formulas $\Gamma \cup \Theta_0 \cup \cdots \cup \Theta_4$
is {\em almost} the required encoding $\Theta_{\myvec{i}}$: the only
problem is that the formulas $\Gamma$ are not in the fragment
$\cN^2$. Massaging them into the appropriate form is the task of the
second stage of the proof.

\vspace{0.25cm}

\noindent
{\bf Second stage: } Recall that, in the first stage, we gave each of
the predicates $X_i$, and $Y_i$ ($0 \leq i < p(n)$), a ``barred''
counterpart $\bar{X}_i$, and $\bar{Y}_i$, with a complementary
interpretation in $\fA$ with respect to $q^\fA = A_1$. Moreover, we
provided a set of formulas $\Theta_0$, guaranteeing that such pairs have
complementary interpretations with respect to the extension of $q$.
Let us do the same for the predicates $X^*_i$, $X^+_{i,j}$,
$X^-_{i,j}$, $Y^*_i$, $Y^+_{i,j}$, $Y^-_{i,j}$ (with indices in the
ranges specified above), letting $\Theta_5$ be the requisite set of
formulas. The construction of $\Theta_5$ is completely routine.

Now enumerate the various predicates $X_i$, $X^*_i$, $X^+_{i,j}$,
$X^-_{i,j}$, $Y_i$, $Y^*_i$, $Y^+_{i,j}$, $Y^-_{i,j}$, in some order, as
\begin{eqnarray}
q_1, \ldots, q_s.
\label{eq:qenumeration}
\end{eqnarray}
(There are indeed $s = 2(p(n)^2 + p(n)+1)$ of these, if you tot them
all up.)  And enumerate their barred counterparts, in the
corresponding order, as
\begin{eqnarray}
\bar{q}_1, \ldots, \bar{q}_s.
\label{eq:qenumerationBar}
\end{eqnarray}
Recall that the `notebook', $A_2$, consists of the elements
$(1,\top), \ldots, (s,\top)$ and $(1,\bot), \ldots,
(s,\bot)$. Referring to the enumerations~\eqref{eq:qenumeration}
and~\eqref{eq:qenumerationBar}, think
of the element $(h,\top)$ as standing for the atom $q_h(x)$, and of
the element $(h,\bot)$ as standing for the atom $\bar{q}_h(x)$, for
all $h$ ($1 \leq h \leq s$). Let $l, l_1, \ldots, l_s$ and $\bar{l}_1,
\ldots, \bar{l}_s$ be new unary predicates, interpreted in $\fA$ as
follows:
\begin{align*}
& l^\fA = A_2 & & \\
& l_h^\fA = \{ (h,\top) \}  & & (1 \leq h \leq s) \\
& \bar{l}_h^\fA = \{ (h,\bot) \} & & (1 \leq h \leq s).
\end{align*}
It is simple to check that $\fA \models \Theta_6$, where $\Theta_6$ is
the set of formulas
\begin{align*}
& \exists_{\leq 2s} x l(x) \\
& \exists x l_h(x) & 
& \exists x \bar{l}_h(x) & & (1 \leq h \leq s) \\
& \forall x (l_h(x) \rightarrow l(x))  & 
& \forall x (\bar{l}_h(x) \rightarrow l(x))  & & (1 \leq h \leq s) \\
& \forall x (l_h(x) \rightarrow \neg l_{h'}(x)) &
& \forall x (\bar{l}_h(x) \rightarrow \neg \bar{l}_{h'}(x)) 
   & & (1 \leq h < h' \leq s) \\ 
& \forall x (l_h(x) \rightarrow \neg \bar{l}_{h'}(x)) 
   & & & & (1 \leq h \leq s, 1 \leq h' \leq s).
\end{align*}
Conversely, in any model of $\Theta_6$, the predicates $l_1, \ldots,
l_s, \bar{l}_1, \ldots, \bar{l}_s$ are uniquely instantiated, and pick
out the $2s$ elements satisfying $l$.  

Fix any formula $\forall x (q(x) \rightarrow \gamma) \in \Gamma$.
Note that the clause $\gamma$ is actually a disjunction of atoms
featuring only the predicates in~\eqref{eq:qenumeration}
and~\eqref{eq:qenumerationBar}.  Let $r_\gamma$ be a new binary
predicate. Since $\fA \models \forall x (q(x) \rightarrow \gamma)$,
define, for each $a \in A_1$, the element $a_\gamma \in A_2$ as
follows. Choose a literal (atom) $L$ of $\gamma$ satisfied by $a$: if
$L$ is $q_h(x)$ for some $h$ ($1 \leq h \leq s$), set $a_\gamma =
\langle h, \top \rangle$; if, on the other hand, $L$ is $\bar{q}_h(x)$
for some $h$ ($1 \leq h \leq s$), set $a_\gamma = \langle h, \bot
\rangle$.  Think of the object $a_\gamma$ as representing some literal
of $\gamma$ satisfied by $a$.  Having defined $a_\gamma$ for all $a
\in A_1$, set
\begin{eqnarray*}
r_\gamma^\fA & = & \{ \langle a,a_\gamma \rangle | a \in A_1 \}.
\end{eqnarray*}
It is then easy to check that $\fA \models \Theta_\gamma$, where
$\Theta_\gamma$ consists of the formula
\begin{align}
& \forall x (q(x) \rightarrow \exists y ( l(y) \wedge r_\gamma(x,y))),
\label{eq:Theta2.1}
\end{align}
together with the following formulas, for all $h$ ($1 \leq h \leq s$):
\begin{align}
& \forall x (q(x) \rightarrow \neg \exists y (l_h(y) \wedge r_\gamma(x,y)))
  & & (\text{if $q_h(x)$ not a literal of $\gamma$})
\label{eq:Theta2.2}\\
& 
\forall x (q(x) \rightarrow \neg \exists y (\bar{l}_h(y) \wedge
r_\gamma(x,y)))
  & & (\text{if $\bar{q}_h(x)$ not a literal of $\gamma$})
\label{eq:Theta2.3}\\
& \forall x (q_h(x) \rightarrow 
      \neg \exists y (\bar{l}_h(y) \wedge r_\gamma(x,y)))
  & & 
\label{eq:Theta2.4}\\
& \forall x (\bar{q}_h(x) \rightarrow 
      \neg \exists y (l_h(y) \wedge r_\gamma(x,y))).
  & & 
\label{eq:Theta2.5}
\end{align}
Conversely, in any model of $\Theta_6 \cup \Theta_\gamma$, \eqref{eq:Theta2.1}
guarantees that every object $a$ in the extension of $q$ is related
via $r_\gamma$ to some object $a_\gamma$ in the extension of $l$
(representing a literal); \eqref{eq:Theta2.2} and~\eqref{eq:Theta2.3}
then state that the literal represented by $a_\gamma$ is a literal of the clause
$\gamma$; and~\eqref{eq:Theta2.4} and~\eqref{eq:Theta2.5} state that
$a$ satisfies this literal. Together, $\Theta_0$, $\Theta_1$,
$\Theta_5$, $\Theta_6$ and $\Theta_\gamma$ thus guarantee that any
object satisfying $q$ also satisfies the clause $\gamma$; in other
words:
\begin{equation}
\Theta_0 \cup \Theta_1 \cup \Theta_5 \cup \Theta_6 
\cup \Theta_\gamma \models \forall x (q(x) \rightarrow \gamma).
\label{eq:theta2good}
\end{equation}
Let $\Theta_7 = \bigcup \{ \Theta_\gamma \mid \mbox{ $\forall x (q(x)
  \rightarrow \gamma) \in \Gamma$} \}$.  

Let $\Theta_{\myvec{i}} = \Theta_0 \cup \cdots \cup \Theta_7$.  The
construction of $\Theta_{\myvec{i}}$ evidently proceeds in time
bounded by a polynomial function of the length $n$ of $\myvec{i}$.
Every formula in $\Theta_{\myvec{i}}$ is an $\cN^2$-formula, modulo
trivial logical manipulations.  And $(C,H,V)$ has a tiling of size $N
= 2^{p(n)}$ with initial segment $\myvec{i}$ if and only if
$\Theta_{\myvec{i}}$ is satisfiable.
\end{proof}
\noindent
We remark that the proof of Theorem~\ref{theo:nt} makes essential use
of binary coding of quantifier subscripts.  For example, the subscript
$MN^2$ has size $\lfloor 2p(n)+ \log M \rfloor  + 1$, and hence
is bounded by a polynomial function of $n$.
\begin{corollary}
The satisfiability problem and finite satisfiability problem for any
logic between $\cN^2$ and $\cC^2$ are both NEXPTIME-complete.
\label{cor:nexp}
\end{corollary}

It follows that determining the validity of arguments in the
numerically definite relational syllogistic is a co-NEXPTIME-complete
problem. Equipping this fragment with relative clauses, for example,
\begin{center}
\begin{minipage}{10cm}
\begin{tabbing}
{\sf At} \= {\sf most 3 artists whom at least 4 beekeepers admire despise at
  least}\\
\> {\sf 5 dentists who envy at most 6 electricians},
\end{tabbing}
\end{minipage}
\end{center}
evidently has no effect on the complexity of determining validity,
since it does not take us outside the fragment $\cC^2$.  Nor 
do proper nouns
or negated
verb-phrases, for example
\begin{center}
\begin{minipage}{10cm}
\begin{tabbing}
{\sf At} \= {\sf most 3 artists do not despise (= fail to despise) at
  least one beekeeper}\\
{\sf At least 3 artists despise Fred}.
\end{tabbing}
\end{minipage}
\end{center}
In fact, we may add a certain amount of anaphora to the fragment while
still remaining within $\cC^2$, thus:
\begin{center}
\begin{minipage}{10cm}
\begin{tabbing}
{\sf At} \= {\sf most 3 artists who despise themselves admire at
  least  4 beekeepers} \\ 
\> {\sf who envy them},
\end{tabbing}
\end{minipage}
\end{center}
though care has to be taken in specifying the precise interpretation
of pronouns (see Pratt-Hartmann~\cite{cndsrs:PH03}).  However, the
complexity-theoretic consequences of extending the repertoire of
quantifiers in $\cC^2$---for example, to include such constructions as
``for most $x$, $\phi$''---are unknown.  

The following related facts
are shown in Pratt-Hartmann and Third~\cite{cndsrs:pht06}: if
sentences involving transitive verbs are added to the ordinary
syllogistic (without numerical quantifiers), the satisfiability
problem for the resulting fragment remains in PTIME; if sentences
involving both transitive verbs and relative clauses are added to the
ordinary syllogistic, the satisfiability problem for the resulting
fragment is EXPTIME-complete.

Although adding relative clauses to $\cN^2$ does not increase the
complexity of satisfiability, it nevertheless has other repercussions of a
logical nature. For one thing, we loose the finite model property: 
the
sentences
\begin{center}
\begin{minipage}{10cm}
  {\sf At least 1 $p$ $r$s at most 0 $p$s}\\
  {\sf At most 0 $p$s are $p$s which at most 0 $p$s $r$}\\
  {\sf At most 0 $p$s $r$ at least 2 $p$s},
\end{minipage}
\end{center}
which, in essence, reproduce the content of formula~\eqref{eq:noFMP},
are satisfiable, but not finitely so.  Hence the satisfiability and
finite satisfiability problems, though both NEXPTIME-complete, are
distinct.  Interestingly, the addition of relative clauses also
affects the question of {\em strong} NEXPTIME-completeness. Inspection of
the proof of Theorem~\ref{theo:nt} shows that binary coding of
quantifier subscripts is required only to overcome the lack of Boolean
connectives in $\cN^2$ (specifically, in simulating the effect of the
formulas $\Gamma$ with the formulas of $\Theta_5$--$\Theta_7$, or
stating that the colours must exhaust the grid). Adding relative
clauses obviates the need for these contortions; and it is in fact
easily checked that the satisfiability problem and the finite
satisfiability problem for this fragment are strongly
NEXPTIME-complete. This difference is 
noteworthy, because some other fragments with counting quantifiers
discussed in the literature have satisfiability and finite satisfiability 
problems whose
complexity is insensitive to whether quantifier subscripts are coded
in unary or binary (Pratt-Hartmann~\cite{cndsrs:ph05,cndsrs:ph07}).
\section{Numerically definite syllogisms}
\label{sec:syllogisms}
Various proof-systems have been proposed in the literature for
determining entailments in the numerically definite syllogistic, based
on numerical generalizations of the traditional syllogisms. Good
examples are the natural deduction systems of
Murphree~\cite{cndsrs:murphree97}, for the language $\cN^{1+}$, and of
Hacker and Parry~\cite{cndsrs:hp67} for the language $\cN^{1}$.  In
this section, we use the results of the foregoing analysis to explore
the possibility of developing a system of numerically definite
syllogisms which is complete, in the sense that all valid sequents
become derivable.

We start by adapting some familiar Aristotelian syllogisms in the
obvious way.  In the sequel, $L$, $L_1$, $L_2$ and $L_3$ 
range over non-ground literals of $\cC^1$---i.e. formulas of the forms
$p(x)$ or $\neg p(x)$. Thus, the formulas of $\cN^{1+}$ simply have the forms
\begin{equation*}
\begin{array}{ll}
\exists_{\geq C} x (L_1 \wedge L_2) \hspace{2cm} &  
\exists_{\leq C} x (L_1 \wedge L_2). 
\end{array}
\end{equation*}
For convenience, we regard $\exists_{\geq C} x (L_1 \wedge L_2)$ and
$\exists_{\geq C} x (L_2 \wedge L_1)$ as identical, and
similarly for their $\exists_{\leq C}$-quantified counterparts.  In
addition, we allow negative numbers to appear in quantifier
subscripts, again with the obvious semantics: for $C < 0$,
$\exists_{\leq C} x (L_1 \wedge L_2)$ is trivially false, and
$\exists_{\geq C} x (L_1 \wedge L_2)$ trivially true.  If $L$ is
a literal, let $\bar{L}$ denote its opposite---that is, the literal
formed by removing any double negation from $\neg L$.  Under these
conventions, define $\cM$ to be the natural deduction system with 
(i) axiom schemas
\begin{equation*}
\exists_{\geq 0} x (L_1 \wedge L_2) 
\hspace{2cm}
\exists_{\leq C} x (L \wedge \bar{L}),
\end{equation*}
for all $C \geq 0$, (ii) rules of inference
\begin{align*}
& \frac{\exists_{\leq C} x (L_1 \wedge L_2) \hspace{1cm} 
      \exists_{\leq D} x (\bar{L}_2 \wedge L_3)}
     {\exists_{\leq (C+D)} x (L_1 \wedge L_3)}
& &
\\
& \frac{\exists_{\geq C} x (L_1 \wedge L_2) \hspace{1cm} 
        \exists_{\leq D} x (L_2 \wedge L_3)} 
     {\exists_{\geq (C-D)} x (L_1 \wedge \bar{L}_3)}
& &
\\
& \frac{\exists_{\leq C} x (L_1 \wedge L_1) \hspace{1cm} 
      \exists_{\geq D} x (L_1 \wedge L_2)}
     {\exists_{\leq (C-D)} x (L_1 \wedge \bar{L}_2)},
& &
\end{align*}
and (iii) the rule of {\em ex falso quodlibet},
allowing the derivation of any formula whatsoever from contradictory
premises:
\begin{quote}
if, from premises $\Phi$, we have deduced the formulas $\exists_{\leq
C} x (L_1 \wedge L_2)$ and $\exists_{\geq D} x (L_1 \wedge L_2)$,
where $D > C$, then we may deduce $\phi$ from $\Phi$.
\end{quote} 

We write $\Phi \vdash_\cM \phi$ if there is a deduction from premises
$\Phi$ to conclusion $\phi$ in $\cM$.  

The system $\cM$ is at least as powerful as that of Murphree, once
notational differences are taken into account.  And Murphree's system
is in turn at least as powerful as that of Hacker and Parry.
Nevertheless, $\cM$ is easily seen not to be complete for the language
$\cN^{1+}$. For example, the valid sequent
\begin{equation}
\exists_{\geq C} x (p(x) \wedge q(x)), 
        \exists_{\geq D} x (p(x) \wedge \neg q(x)) \models
\exists_{\geq (C+D)} x (p(x) \wedge p(x))
\label{eq:sequent}
\end{equation}
is not derivable in $\cM$. (Possibly, these writers never intended
their systems to handle conclusions of this form.) The question
therefore arises as to the prospects for producing a complete system
of syllogisms for the numerically definite syllogistic.

To make the ensuing analysis more robust (and the comparison with
published systems fairer), we consider the special case of the
validity problem in which it is known how many objects satisfy each
predicate in question, and how many objects fail to do so.  Formally,
we consider only inference problems from {\em numerically explicit}
premise sets, in the following sense.
\begin{definition}
Let $\Phi$ be a set of $\cN^{1+}$-formulas, and let $p_1, \ldots, p_n$
be the predicates appearing in $\Phi$.  We say that $\Phi$ is {\em
numerically explicit} if there exist natural numbers $C$, $C_1,
\ldots, C_n$ with $C > 0$ such that, for all $i$ ($1 \leq i \leq n$),
(i) $C_i \leq C$, and (ii) $\Phi$ contains the formulas
\begin{equation*}
\begin{array}{lll}
\exists_{\leq C_i} x (p_i(x) \wedge p_i(x)) & &
\exists_{\leq (C - C_i)} x (\neg p_i(x)\wedge \neg p_i(x))\\
\exists_{\geq C_i} x (p_i(x)  \wedge p_i(x)) & & 
\exists_{\geq (C - C_i)} x (\neg p_i(x)\wedge \neg p_i(x)).
\end{array}
\end{equation*}
\label{def:csep}
\end{definition}
\noindent
Regarding the sequent~\eqref{eq:sequent}, it is easy to show that, 
if $\Phi$ is any
numerically explicit premise set containing $\exists_{\geq C} x (p(x)
\wedge q(x))$ and $\exists_{\geq D} x (p(x) \wedge \neg q(x))$, then
$\Phi \vdash_\cM \exists_{\geq (C+D)} x (p(x) \wedge p(x))$.
We remark in passing that de Morgan's numerically
definite syllogisms also make reference to the assumed cardinalities
of some of the terms they involve (de Morgan~\cite{cndsrs:deMorgan47},
p.~161).

Unfortunately, the prospects for a complete system of numerical
syllogisms, even for the special case of numerically explicit premise
sets, are not bright. For, in the sequel, we exhibit a numerically
explicit set of $\cN^{1+}$-formulas $\Phi$ and an $\cN^{1+}$-formula
$\phi$ such that $\Phi \models \phi$, but $\Phi \not \vdash_\cM \phi$.
Moreover, this incompleteness result will be seen to be relatively
robust under a range of conceivable extensions of $\cM$.

For readability, we shall henceforth contract $\cN^{1+}$-formulas with
repeated literals: thus, $\exists_{\leq C} x(p(x) \wedge p(x))$
becomes $\exists_{\leq C} x p(x)$, etc. We use the quantifier
$\exists_{=C}$ to abbreviate the obvious pair of formulas involving
$\exists_{\leq C}$ and $\exists_{\geq C}$. And we write
$\cN^1$-formulas of the form $\exists_{\leq 0} x(p(x) \wedge \pm
q(x))$ in their more familiar guise: $\forall x(p(x) \rightarrow \mp
q(x))$.  Fix $m \geq 6$. Let $\mymat{A}$ and $\myvec{c}$ be as in
Lemma~\ref{lma:manyNonZeros}, and let $\Phi_1$ be the set of
$\cN^1$-formulas consisting of
\begin{align}
& \exists_{\leq 3(m+1)} x t(x)
  & & 
\label{eq:Phi1.4}\\
& \exists_{\geq 3} x t_j(x)
  & & (1 \leq j \leq m+1)
\label{eq:Phi1.1}\\
& \forall x (t_j(x) \rightarrow t(x))
  & & (1 \leq j \leq m+1)
\label{eq:Phi1.2}\\
& \forall x (t_j(x) \rightarrow \neg t_{j'}(x))
  & & (1 \leq j < j'\leq m+1)
\label{eq:Phi1.3}\\
& \forall x (s_i(x) \rightarrow t(x))
  & & (1 \leq i \leq m)
\label{eq:Phi2.0}\\
& \forall x (t_j(x) \rightarrow s_i(x))
  & & (1 \leq i \leq m, 1 \leq j \leq m+1, \myvec{A}_{i,j} = 1)
\label{eq:Phi2.1}\\
& \forall x (t_j(x) \rightarrow \neg s_i(x))
  & & (1 \leq i \leq m, 1 \leq j \leq m+1, \myvec{A}_{i,j} = 0)
\label{eq:Phi2.2}\\
& \exists_{=3}x (s_i(x) \wedge r(x))
  & & (1 \leq i \leq m-1)
\label{eq:Phi3.1}\\
& \exists_{=4}x (s_m(x) \wedge r(x)).
\label{eq:Phi3.2}
\end{align}
Note that the list of quantifier subscripts in the $m$ formulas
of~\eqref{eq:Phi3.1} and~\eqref{eq:Phi3.2} matches the vector
$\myvec{c}$.
\begin{claim}
For all $j$ $($$1 \leq j \leq m+1$$)$, $\Phi_1 \models \exists_{\geq 1}
x (t_j(x) \wedge r(x))$.
\label{claim:valid}
\end{claim}
\begin{proof}
Suppose $\fA \models \Phi_1$. From~\eqref{eq:Phi1.4}--\eqref{eq:Phi1.3},
$t^\fA$ is partitioned into the pairwise disjoint sets
$t_1^\fA$, \ldots, $t_{m+1}^\fA$. 
And so, from~\eqref{eq:Phi2.0}--\eqref{eq:Phi2.2}, we have,
for all $i$ ($1 \leq i \leq m$),
\begin{equation*}
s_i^\fA = \bigcup \{t_j^\fA \mid \mymat{A}_{i,j} = 1 \},
\end{equation*}
and hence
\begin{equation*}
|s_i^\fA \cap r^\fA|  = 
   \sum \{ |t_j^\fA \cap r^\fA|  : \mymat{A}_{i,j} = 1 \}.
\end{equation*}
Therefore, \eqref{eq:Phi3.1} implies, for all $i$ ($1 \leq i \leq m-1$), 
\begin{equation*}
   \sum \{| t_j^\fA \cap r^\fA | : \mymat{A}_{i,j} = 1 \} = 3,
\end{equation*}
while~\eqref{eq:Phi3.2} implies
\begin{equation*}
   \sum \{| t_j^\fA \cap r^\fA | : \mymat{A}_{m,j} = 1 \} = 4.
\end{equation*}
In other words, $(| t_1^\fA \cap r^\fA |, \ldots, | t_{m+1}^\fA \cap r^\fA|)^T$
is a solution of $\mymat{A}\myvec{x} = \myvec{c}$.
Applying Lemma~\ref{lma:manyNonZeros}, $| t_j^\fA \cap r^\fA | = 1$
for all $j$ ($1 \leq j \leq m+1$), which proves the claim.
\end{proof}

\begin{figure}
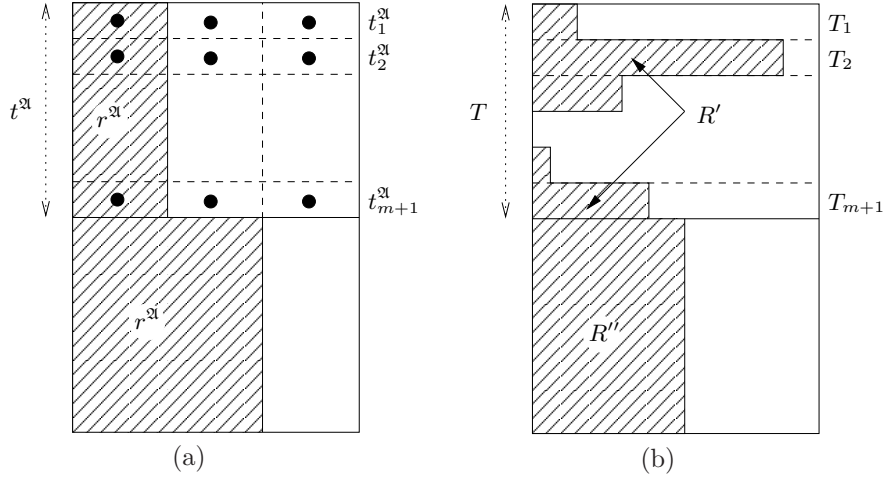

\begin{minipage}{5cm}
\begin{center}
\input{modelOrd.pstex_t}\\
(a)
\end{center}
\end{minipage}
\hspace{1cm}
\begin{minipage}{5cm}
\begin{center}
\input{modelProb.pstex_t}\\
(b)
\end{center}
\end{minipage}
\caption{Models of $\Phi$: (a) standard semantics; (b) probabilistic
  semantics. In (b), one of the sets $R_j = R' \cap T_j$ is empty.}
\label{fig:model}
\end{figure}

Now let $\Phi_2$ be the set of $\cN^{1+}$-formulas
\begin{equation*}
\begin{array}{lllll}
\exists_{= 3(m+1)} x t(x) & \exists_{= 3(m+1)} x \neg t(x) \\    
\exists_{= 3} x t_j(x) & \exists_{= 6m+3} x \neg t_j(x)
  & & & (1 \leq j \leq m+1)  \\
\exists_{= 9} x s_i(x) & \exists_{= 6m - 3} x \neg s_i(x)
  & & & (1 \leq i \leq m-1) \\
\exists_{= 12} x s_m(x) & \exists_{= 6m -6} x \neg s_m(x)\\
\exists_{= 3(m+1)} x r(x) & \exists_{= 3(m+1)} x \neg r(x),
\end{array}
\end{equation*}
and let $\Phi = \Phi_1 \cup \Phi_2$.  Thus, $\Phi$ is numerically
explicit.  Moreover, $\Phi$ is satisfiable: Fig.~\ref{fig:model}a
depicts a (in fact, {\em the}) model $\fA$ of $\Phi$. The domain $A$
has cardinality $6(m+1)$, equally split between $t^\fA$ and its
complement; the sets $t_j^\fA$ ($1 \leq j \leq m+1$) partition $t^\fA$
into 3-element sets; the set $r^\fA$ has cardinality $3(m+1)$; and the
sets $t_j^\fA \cap r^\fA$ are all singletons. The extensions of the
$s_i$ (not indicated in Fig.~\ref{fig:model}a, for clarity) are all
unions of various $t_j^\fA$, as specified by the matrix $\mymat{A}$.
Nevertheless, the validities reported in Claim~\ref{claim:valid}
cannot all be reproduced by the proof-system $\cM$.
\begin{claim}
There exists a $j$ $($$1 \leq j \leq m+1$$)$ such that
$\Phi {\not \vdash}_\cM \exists_{\geq 1} x (t_j(x) \wedge r(x))$.
\label{claim:underivable}
\end{claim}
\begin{proof}
Fix $N = 6(m+1)$.  We give a completely new semantics for the language
$\cN^{1+}$ as follows. Let $\Sigma$ be the (assumed countable) set of
all unary predicates available to $\cN^{1+}$.  Let us re-badge the
elements $p_1, p_2, \ldots$ of $\Sigma$ as proposition letters; and,
as before, we let $\cS$ denote the propositional language over this
signature. If $P$ is a probability assignment for $\cS$, we
interpret $\cN^{1+}$ with respect to $P$ by writing
\begin{equation*}
P \kludgeSem \exists_{\geq C} x(p_1(x) \wedge p_2(x)) 
\text{ if and only if }  P(p_1 \wedge p_2) \geq C/N,
\end{equation*}
and similarly for all the other forms of $\cN^{1+}$. (Remember that
$N$ is a constant here.)  It is readily verified that the proof-system
$\cM$ is sound for the \kludgeSem -semantics: all instances of the 
axiom schemas are true; all instances of the three inference rules are
truth-preserving; and {\em ex falso quodlibet} is
validity-preserving. We proceed to construct a probability assignment
$P$ such that: (i) $P \kludgeSem \Phi$, and (ii) for some
$j$ ($1 \leq j \leq m+1$), $P \not \kludgeSem \exists_{\geq 1} x
(t_j(x) \wedge r(x))$. It follows that, for this $j$, $\Phi
{\not \vdash}_\cM \exists_{\geq 1} x (t_j(x) \wedge r(x))$.

By Lemma~\ref{lma:fewNonZeros3}, the equations $\mymat{A}\myvec{x} =
\myvec{c}$ have a solution $u_1, \ldots, u_{m+1}$ over $\Q^+$ with at
least one zero value. On the other hand, it is obvious from
examination of $\mymat{A}$ and $\myvec{c}$ that $u_j$ must be less
than or equal to 3 for all $j$ ($1 \leq j \leq m+1$). Let $u$ be the
least common multiple of all the (non-zero) denominators in the $u_j$;
let $W$ be a set of $uN = 6u(m+1)$ objects (henceforth: ``worlds'');
and let $T$ be a subset of $W$ of cardinality $3u(m+1)$. Now let $T$
be partitioned into cells $T_1, \ldots, T_{m+1}$, each of which
contains $3u$ worlds.  For each $j$ ($1 \leq j \leq m+1$), let $R_j$
be a subset of $T_j$ of cardinality $uu_j$ (which must be a natural
number no greater than $3u$), and let $R' = \bigcup \{R_j \mid 1 \leq
j \leq m+1 \}$.  Since $R' \subseteq T$, $|T| = 3u(m+1)$, and $|W| =
6u(m+1)$, we may choose a set $R'' \subseteq W \setminus T$ such that
the set $R = R' \cup R''$ has cardinality 3u(m+1).  Finally, for each
$i$ ($1 \leq i \leq m$), let $S_i = \bigcup \{T_j \mid \mymat{A}_{i,j}
= 1 \}$. Thus, $S_i$ has cardinality $9u$ for all $i$ ($1 \leq i \leq
m-1$), and $S_m$ has cardinality $12u$. This arrangement is depicted,
schematically, in Fig.~\ref{fig:model}b, except that the $S_i$ are not
indicated, for clarity. Note, however, that, because $u_1, \ldots,
u_{m+1}$ is a solution of $\mymat{A}\myvec{x} = \myvec{c}$, $|S_i \cap
R | = 3u$, for all $i$ ($1 \leq i \leq m-1$), and $|S_m \cap R | =
4u$.

We associate with each world $w \in W$ a truth-value assignment
$\theta_w$ for the propositional language $\cS$ 
by setting $\theta_w(s_i) = \top$ if and only if $w \in
S_i$, $\theta_w(t_j) = \top$ if and only if $w \in T_j$, $\theta_w(t)
= \top$ if and only if $w \in T$, $\theta_w(r) = \top$ if and only if
$w \in R$, and $\theta_w(p) = \bot$ for all other proposition-letters
$p$. Then we define the probability assignment $P$ by taking a flat
distribution on $W$. That is, for every $\phi \in \cS$, set
\begin{equation*}
P(\phi) = | \{ w \in W:  \theta_w \models \phi \} |/(uN).
\end{equation*}
It is simple to check that $P \kludgeSem \Phi$. On the other
hand, at least one of the $u_j$ is zero; and for this value of $j$,
$P(t_j \wedge r) = 0$, whence $P \not \kludgeSem \exists_{\geq 1} x
(t_j(x) \wedge r(x))$.
\end{proof}
Hence we have:
\begin{theorem}
The proof-system $\cM$ is not complete, even for numerically explicit
sets of premises.
\label{theo:M}
\end{theorem}
Theorem~\ref{theo:M} is robust with respect to {\em any} strenthening
of $\cM$ that is sound under the probabilistic interpretation in the
proof of Claim~\ref{claim:underivable}. We mention that, in another
paper, Murphree presents a language similar to our $\cN^2$
(Murphree~\cite{cndsrs:murphree98}); however, no systematic proof
theory is developed.  In fact, we are not aware of any published
system of numerically definite syllogisms which has been shown to be
complete.  
\bibliographystyle{plain} \bibliography{cndsrs}
\end{document}